\documentclass[]{bytedance_seed}

\usepackage[toc,page,header]{appendix}

\usepackage{xcolor}      
\usepackage{amssymb}     
\usepackage{pifont}
\usepackage{todonotes}
\usepackage{minitoc}
\usepackage{threeparttable} 
\usepackage{stackengine} 
\usepackage{graphicx}    
\usepackage{fdsymbol}   
\usepackage{longtable}
\usepackage{multirow}
\usepackage{booktabs}
\usepackage{array}
\usepackage{pdflscape} 
\usepackage{rotating}
\usepackage{colortbl}

\newcommand{\greencheck}{{\color{green!70!black}{\ding{52}}}}   
\newcommand{\redcross}  {{\color{red!85!black} {\ding{56}}}}    
\newcommand{\redstar}  {{\color{red!85!black} {\ding{72}}}}

\title{AetherCode: Evaluating LLMs' Ability to Win In Premier Programming Competitions}

\author[]{ByteDance, M-A-P}

\affiliation[]{Full author list in Contributions}

\abstract{
Competitive programming has emerged as a critical benchmark for evaluating the reasoning and coding capabilities of Large Language Models (LLMs). Despite impressive progress on existing benchmarks, we argue that current evaluations overstate model proficiency, masking a substantial gap between LLMs and elite human programmers. This gap arises from two key limitations: insufficient difficulty and scope of benchmark problems, and evaluation bias from low-quality test cases. To address these shortcomings, we present \textbf{AetherCode}, a new benchmark that draws problems from premier programming competitions such as IOI and ICPC, offering broader coverage and higher difficulty. AetherCode further incorporates comprehensive, expert-validated test suites built through a hybrid of automated generation and human curation, ensuring rigorous and reliable assessment. By combining challenging problem design with robust evaluation, AetherCode provides a more faithful measure of LLM capabilities and sets a new standard for future research in code reasoning.
}

\date{\today}
\correspondence{Zihan Wang at \email{zh.wang@bytedance.com}, Jiaze Chen at \email{chenjiaze@bytedance.com}}

\checkdata[Project Page]{\url{https://huggingface.co/datasets/m-a-p/AetherCode}}

\begin{document}
\maketitle


\section{Introduction}

Competitive programming is widely regarded as a crucial benchmark for evaluating the reasoning and coding capabilities of Large Language Models (LLMs) \cite{o1}. Solving complex competitive programming problems demands not only sophisticated reasoning abilities but also knowledge from diverse domains, including mathematics, data structures, and algorithms. Recent years have witnessed rapid advancements in the reasoning capabilities of LLMs, a key indicator of which is their success on a majority of existing code reasoning benchmarks. State-of-the-art models now achieve over 90\% \textit{Pass@1} accuracy on MBPP \cite{mbpp} and HumanEval \cite{humaneval}, and over 80\% on LiveCodeBench \cite{livecodebench}. These encouraging developments might lead one to ask: has competitive programming been mastered by LLMs?

In this paper, we argue that a significant gap still exists between the performance of LLMs and top-tier human competitors in programming contests. We propose that the perception of LLM dominance stems primarily from the limitations in the breadth and rigor of current code reasoning benchmarks, which are no longer sufficient to fully assess the capabilities of today's increasingly powerful models. Specifically, we identify two main shortcomings in existing benchmarks:

\begin{itemize}
    \item \textbf{Insufficient Difficulty and Scope.} Early benchmarks such as HumanEval \cite{humaneval} and MBPP \cite{mbpp} consist of basic coding tasks, for instance, sorting or reversing a list, which present minimal reasoning challenges for state-of-the-art LLMs. More recent ``competition-level'' benchmarks often source problems from a limited set of websites. For example, LiveCodeBench \cite{livecodebench} collects problems mainly from LeetCode and AtCoder, while CodeELO \cite{codeelo} and LiveCodeBench Pro \cite{lcbpro} focus solely on CodeForces. Each of these websites has inherent limitations. LeetCode problems are generally easier and often require only the implementation of a single function rather than a complete program. CodeForces contests, which typically feature 5-7 problems within a 2-3 hour timeframe, constrain the design space for problem setters, for example, leading to a scarcity of problems that require complex, large-scale implementations.
    \item \textbf{Evaluation Bias from Low-Quality Test Cases.} The correctness of a piece of code is verified using a comprehensive set of test cases (input-output pairs). An incomplete test suite may fail to detect incorrect submissions, particularly those with subtle flaws, such as the mishandling of corner cases or solutions that exceed time limits under specific, extreme conditions. Consequently, designing high-quality test cases is a huge challenge that requires a deep understanding of potential failure points, a skill typically honed through extensive competitive programming experience. Most past benchmarks lack sufficiently rigorous test cases. HumanEval \cite{humaneval} and MBPP \cite{mbpp}, for instance, rely on a small number of handwritten test cases. Others, including EvalPlus \cite{evalplus}, CodeContests \cite{codecontests}, and LiveCodeBench \cite{livecodebench}, employ naive test case generation pipelines, such as random mutation, which fall far short of the quality of expert-designed test suites. Furthermore, recent research \cite{gvc} has revealed issues with test case correctness itself; for example, many test cases in the CodeContests dataset do not adhere to the problem's constraints, causing even correct solutions to fail. It is worth noting that some recent benchmarks, such as CodeELO \cite{codeelo} and LiveCodeBench Pro \cite{lcbpro}, have attempted to leverage the official CodeForces judging service to indirectly access its high-quality, expert-crafted test cases. However, this approach presents two significant issues. First, it raises compliance risks, as CodeForces explicitly prohibits the use of crawlers on its judging interface. Second, this method is constrained by submission frequency limits, which impedes agile and flexible experimentation. Therefore, we contend that an open-source benchmark with high-quality, self-contained test cases remains critically important for the LLM community.

\end{itemize}

To address these challenges, we introduce AetherCode, a new benchmark with the following key contributions:

\textbf{Problem Curation from Top-Tier Competitions.} AetherCode is the first benchmark to systematically collect problems from premier programming competitions worldwide, including the Olympiad in Informatics (OI) and the International Collegiate Programming Contest (ICPC). Our process involved a comprehensive collection, meticulous cleaning, and format conversion of problems from PDF to a Markdown+LaTeX structure. Each problem statement was manually proofread for correctness, and a team of competitive programming experts annotated each problem with classification tags.

\textbf{High-Quality Test Case Generation.} We developed a hybrid methodology, combining automated generation with expert annotation, to create high-quality test cases for every problem. We evaluated the correctness and comprehensiveness of our test cases by validating them against a large corpus of collected solutions, enforcing a standard of zero false positives and zero false negatives.

This paper is organized as follows: Section \ref{sec:curation} details the benchmark curation process. Section \ref{sec:eval} presents our evaluation results. Section \ref{sec:conclusion} concludes the paper and discusses directions for future research.

\begin{table*}[t]
    \footnotesize
    \centering
    \renewcommand\arraystretch{1.2}
     \resizebox{\linewidth}{!}{

\begin{threeparttable}
     \caption{
 Comparison between AetherCode and other code reasoning benchmarks
}
\begin{tabular}{@{}lccccc@{}}
\toprule
\small
\textbf{Dataset}   & \textbf{Difficulty} & \textbf{\# Problems} &\textbf{Updates}  & \textbf{\parbox{2cm}{\centering Test Cases \\ Construction}}  & \textbf{Source} \\
\midrule
HumanEval \citep{humaneval} & \redstar & 164 &\redcross & Handcrafted & Original\\
MBPP \citep{mbpp} & \redstar & 974 &\redcross & Handcrafted &Original\\
APPS \cite{apps} & \redstar\redstar\redstar & 5,000 & \redcross & Crawled & CodeForces, AtCoder~\textit{etc.}\\
USACO \citep{usaco} & \redstar\redstar\redstar & 307 &\redcross &Publicly accessible & USACO\\
CodeContests \cite{codecontests} & \redstar\redstar\redstar & 165 & \redcross & Mutation & CodeForces, AtCoder~\textit{etc.}\\
LiveCodeBench \cite{livecodebench} & \redstar\redstar & 1055 &\greencheck & Semi-automatic & LeetCode, AtCoder\\
CodeELO \cite{codeelo} & \redstar\redstar\redstar & 387 &\greencheck & \redcross &CodeForces\\
LiveCodeBench Pro \cite{lcbpro} & \redstar\redstar\redstar & 584 &\greencheck & \redcross &CodeForces\\
\midrule
AetherCode & \redstar\redstar\redstar\redstar & 456 &\greencheck & G-V Agent\cite{gvc} \& Experts & Premier Contests\\

\bottomrule
\end{tabular}
\end{threeparttable}
}
\label{tab:comparison}
\end{table*}
\section{Benchmark Curation}
\label{sec:curation}

This Section details the curation process of the AetherCode Benchmark. Sections \ref{sec:problem_collection} and \ref{sec:problem_categorizing} describe the specifics of problem collection and categorizing, respectively. Section \ref{sec:test_case} explains how we construct high-quality test cases for each problem, and Section \ref{sec:stats} presents the statistical data of AetherCode.

\subsection{Problem Collection} 
\label{sec:problem_collection}

We source our problems from premier programming competitions worldwide rather than from online programming websites. Based on their target audience, these competitions can be broadly categorized into two main series: the Olympiad in Informatics (OI) series, which is aimed at pre-college school students, and the International Collegiate Programming Contest (ICPC) series, which is designed for college students.

\textbf{OI Series.} The Olympiad in Informatics is a series of competitions aimed at popularizing computer science knowledge among middle-school students and cultivating outstanding talents in computer science. The OI competitions usually require participants to solve algorithm-related problems by programming. Take the International Olympiad in Informatics (IOI), the top-level event of OI, as an example. Each contestant competes individually, and each country can send up to 4 players. During the two-day competition, players need to independently solve 3 problems within 5 hours each day, mainly using C++. Furthermore, various countries and regions host their own national or regional OI competitions, such as the National Olympiad in Informatics (NOI) in China and the USA Computing Olympiad (USACO) in the United States. Top-performing contestants in these competitions earn the opportunity to advance to the IOI.

\textbf{ICPC Series.} The ICPC is the oldest, largest, and most prestigious university-level programming contest in the world. Each team consists of up to 3 students and uses one computer to solve 10 - 13 problems in 5 hours, using programming languages such as C, C++, Java, or Python. The team that correctly solves the most problems with the least total time wins.

The world is divided into several regions for the ICPC. In Europe, there are Central Europe (CERC), North Europe (NWERC), South-East Europe (SEERC), and South-West Europe (SWERC) regions. Other regions include Asia-Pacific, Asia East Continent, North America, Latin America, Africa, and Arab region, etc. The ICPC is a multi-tiered event. First, there are \textbf{regional contests} held worldwide from September to November each year. The top-performing teams in the regional contests advance to the \textbf{regional finals or championships}. Then, the best teams from these finals or championships qualify for the ICPC \textbf{World Finals}, which is usually held from April to June each year. This is the highest-level stage of the ICPC, where the best teams from around the world compete for the championship.

In addition to the official ICPC events, we also incorporated problems from other large-scale and renowned collegiate programming contests, such as the China Collegiate Programming Contest (CCPC).

For each problem, we collected the following components:

\begin{itemize}
    \item \textbf{Problem Statement.} The statement typically comprises a title, a detailed problem description, input/output specifications, sample inputs and outputs with explanations, data range constraints, and time/memory limits. The majority of the problem statements was originally in PDF format. To enhance comprehension for LLMs, we converted these PDFs into a Markdown format with LaTeX for mathematical notations. Each converted file was then manually proofread to ensure its accuracy.
    \item \textbf{Solutions.} We curated a collection of over 30,000 human-written solutions for these problems, encompassing both correct and incorrect submissions. For each problem, we ensured a minimum of 5 correct and 20 incorrect solutions. The primary purpose of collecting these solutions is to evaluate the quality of the subsequently generated test cases, a process detailed in Section \ref{sec:test_case}.
    \item \textbf{Test Cases.} A minority of the competitions, e.g., USACO, publicly released their official test cases, which we collected and standardized. For problems where official test cases were not available, we constructed our high-quality test cases. The methodology for this construction is described in Section \ref{sec:test_case}.
    \item \textbf{Metadata.} We also gathered auxiliary information, such as the date of the competition (for decontamination purposes) and human contestant performance data (to facilitate difficulty assessment), among other available data points.
\end{itemize}

\subsection{Problem Categorization}
\label{sec:problem_categorizing}

Beyond curating problems, an equally critical step in constructing AetherCode was the systematic categorization of each problem to ensure comprehensive coverage and facilitate fine-grained evaluation. To this end, we adopted a multi-dimensional categorization framework designed with the input of competitive programming experts:

\begin{enumerate}
    \item \textbf{Difficulty Segmentation.} Problems were divided into four levels of difficulty: \textit{Easy}, \textit{Medium}, \textit{Hard}, and \textit{Extreme}. This classification was guided by expert judgment as well as official contest results. Notably, problems that no human contestant was able to solve during a competition were classified as \textit{Extreme}, representing challenges that push the boundaries of algorithmic reasoning.
    
    \item \textbf{Temporal and Contextual Dimensions.} Each problem was annotated with metadata to enable both decontamination and longitudinal analysis of model performance:
    \begin{itemize}
        \item Year of the contest, allowing chronological tracking of trends in problem design and model capabilities.
        \item Competition type, primarily distinguishing between Olympiad in Informatics (OI) and International Collegiate Programming Contest (ICPC) series. 
        \item Competition scope, categorizing contests as regional-level, national-level, continental-level, or world finals. 
    \end{itemize}
    
    \item \textbf{Problem Format Constraints.} Some problems require additional considerations beyond a standard input–output interface:
    \begin{itemize}
        \item Problems dependent on visual or image-based input were excluded from the benchmark. 
        \item Problems requiring special judges or custom checkers were explicitly labeled to ensure proper handling during evaluation.
    \end{itemize}
    
    \item \textbf{Algorithmic and Domain Categories.} To capture the breadth of algorithmic knowledge tested in programming contests, we implemented a hierarchical taxonomy as shown in Table~\ref{tab:algorithm_categories}:
    \begin{itemize}
        \item Primary categories correspond to major domains such as \textit{Dynamic Programming}, \textit{Graph Theory}, \textit{Computational Geometry}, \textit{Data Structures}, and \textit{Mathematics}.
        \item Secondary categories provide finer granularity, such as \textit{tree dynamic programming}, \textit{flow algorithms}, \textit{convex hull geometry}, or \textit{modular arithmetic}. Problems can belong to multiple categories to reflect their cross-disciplinary nature.
    \end{itemize}
\end{enumerate}

This structured categorization enables targeted evaluation of model strengths and weaknesses while also ensuring that AetherCode serves as a scalable resource for future research. In particular, it allows progress to be tracked across difficulty levels, problem types, and algorithmic domains, providing a more comprehensive understanding of model capabilities.


\subsection{Test Case Construction}
\label{sec:test_case}

Recent studies \cite{evalplus, gvc} have highlighted concerns regarding the quality of test cases in several existing code datasets. For instance, benchmarks such as MBPP \cite{mbpp} and HumanEval \cite{humaneval} include only a limited number of handwritten test cases per problem. Others, like CodeContests \cite{codecontests} and EvalPlus \cite{evalplus}, rely on naive methods such as mutation to generate test cases. Consequently, such test cases are insufficient for comprehensively evaluating the correctness and efficiency of a program. Therefore, we contend that the quality of test case construction is a critical factor determining the overall quality of a benchmark.

Notably, some recent benchmarks \cite{codeelo, lcbpro} directly utilize the CodeForces's judging service for evaluation. This approach allows them to indirectly access high-quality test cases created by professional problem setters, thereby circumventing the challenge of test case construction. However, this method presents potential compliance risks, as CodeForces explicitly prohibits the use of crawlers on its judging interface. Furthermore, this approach is constrained by submission frequency limits, which impedes agile and flexible evaluation. Therefore, we argue that a benchmark equipped with its own high-quality test cases remains critically important for the LLM community.

To ensure AetherCode possesses sufficiently high-quality test cases, we approached the task from two perspectives. First, we established more stringent evaluation criteria for test case quality, which is presented in Section \ref{sec:test_case_qa}. Second, we employed a hybrid approach, combining automated generation with expert annotation, to construct the test cases, which are presented in Sections \ref{sec:auto_test_case} and \ref{sec:human_test_case}.

\subsubsection{Test Case Quality Assessment}
\label{sec:test_case_qa}

Previous research on test case quality has predominantly focused on quantity, operating under the assumption that a greater number of test cases correlates with higher quality \cite{codecontests, li_taco_2023}. However, recent studies \cite{gvc} indicate that quantity is not a direct proxy for quality. This discrepancy arises from two primary issues. First, test cases in some older datasets, despite their volume, suffer from significant correctness issues, often violating the problem's explicit constraints. Second, conventional test case generation methods that merely amass large volumes of random data fail to provide adequate coverage of various special and corner cases.

Consequently, we depart from evaluating test cases by their quantity and instead propose a direct assessment of their ability to discriminate between correct and incorrect solutions. In our framework, we conceptualize the entire test suite for a problem as a binary classifier, that is, a classifier that distinguishes between correct and incorrect solutions. We then evaluate the performance of this classifier using a large, curated collection of both correct and incorrect submissions. We adopt the True Positive Rate (TPR) and True Negative Rate (TNR) as our primary evaluation metrics.

\begin{align}
    \mathrm{TPR}&=\frac{\mathrm{True\ Positive}}{\mathrm{True\ Positive}+\mathrm{False\ Negative}}=\frac{\mathrm{Number\ of\ Passed\ Correct\ Solutions}}{\mathrm{Number\ of\ Correct\ Solutions}}\\
    \mathrm{TNR}&=\frac{\mathrm{True\ Negative}}{\mathrm{True\ Negative}+\mathrm{False\ Positive}}=\frac{\mathrm{Number\ of\ Rejected\ Incorrect\ Solutions}}{\mathrm{Number\ of\ Incorrect\ Solutions}}
\end{align}

The TPR measures the \textbf{correctness} of the test cases; a high TPR indicates that correct solutions are not erroneously failed, which is expected when the test cases themselves are valid. Conversely, the TNR measures the \textbf{comprehensiveness} or \textbf{coverage} of the test cases, quantifying their ability to detect (or ``hack'') incorrect solutions.

By employing a hybrid approach that combines automated generation with expert curation, we have achieved a 100\% TPR and 100\% TNR on our collected solution set. This signifies that all collected correct solutions pass our test cases, while all collected incorrect solutions are successfully rejected. To the best of our knowledge, AetherCode is the first benchmark that sets such a high standard for test cases.

\subsubsection{Automatic Construction of Test Cases}
\label{sec:auto_test_case}

We employed the Generator-Validator Agent System \cite{gvc} to automatically construct the test cases, a methodology whose effectiveness has been well-established in prior research \cite{gvc}. Building upon this foundation, we incorporated an additional step of manual verification for the Validator. This step ensures the Validator's correctness, thereby guaranteeing that all generated test cases adhere to every constraint specified in the problem description.

Recognizing that the initial Automatic Construction phase could not achieve a 100\% TNR on its own, we introduced an additional expert annotation stage to further strengthen the test cases.

\subsubsection{Expert Annotation of Test Cases}
\label{sec:human_test_case}

To this end, we recruited 67 competitive programming experts. The majority of them hold Codeforces ratings above 2000, with one expert exceeding 2600 and achieving the title of International Grandmaster. These experts were tasked with constructing targeted test cases specifically designed to fail the various incorrect solutions we had collected. These manually crafted test cases were then merged with the automatically generated ones to form the final test suite.

Furthermore, we recognized that for certain problems with a limited number of collected incorrect solutions (fewer than 50), achieving a 100\% TNR might not sufficiently guarantee the robustness of the test cases. To address this, we subjected the test cases for all problems to a manual quality audit by a specialized review team. Each member of this elite team holds at least three ICPC gold medals and has a minimum of two years of experience in competitive programming problem-setting. Their deep understanding of potential pitfalls and common errors in each problem allows them to leverage their extensive experience to further ensure the quality and comprehensiveness of the test cases.

Additionally, for problems that accept multiple valid outputs, customized judging scripts (a.k.a. checker, or special judge) were provided and thoroughly reviewed by these experts to ensure correct evaluation.

\subsection{Statistics}
\label{sec:stats}

The number of problems in different difficulties and years of AetherCode v1 is presented in Table \ref{tab:difficulty_year}. The number of problems in different categories of AetherCode v1 is presented in Table \ref{tab:category}.

\begin{table}[htbp]
\centering
\caption{The number of problems in different difficulties and years of AetherCode v1 (2401-2505).}
\label{tab:difficulty_year}
\begin{tabular}{cccc|cc}
\toprule
\multicolumn{4}{c|}{\textbf{Difficulty}} & \multicolumn{2}{c}{\textbf{Year}} \\
\midrule
Easy & Medium & Hard & Extreme & 2024 & 2025 \\
\midrule
159 & 145 & 132 & 20 & 400 & 56 \\
\bottomrule
\end{tabular}
\end{table}
\begin{table}[htbp]
\centering
\caption{The number of problems in different categories of AetherCode v1 (2401-2505).}
\label{tab:category}
\begin{tabular}{l|cccccccccc}
\toprule
\textbf{Category} & \textbf{Basic} & \textbf{Search} & \textbf{DP} & \textbf{Str.} & \textbf{Math} & \textbf{DS} & \textbf{Graph} & \textbf{Geo.} & \textbf{Tech.} & \textbf{Tree} \\
\midrule
Count & 225 & 50 & 110 & 26 & 96 & 120 & 64 & 36 & 147 & 24 \\
\bottomrule
\end{tabular}
\end{table}


\section{Evaluation}
\label{sec:eval}
Our evaluation includes 8 reasoning models and 5 non-reasoning models. The reasoning models comprise \texttt{o4-mini-high}~\citep{OpenAI_o4mini}, \texttt{Gemini-2.5-Pro/Flash}~\citep{Google}, \texttt{Seed-1.6-Thinking}~\citep{ByteDance}, \texttt{DeepSeek-R1}~\cite{guo2025deepseek}, and \texttt{Qwen3}~\citep{yang2025qwen3}, among others. The non-reasoning models consist of \texttt{GPT-4.1}~\cite{OpenAI_GPT41}, \texttt{GPT-4o}~\citep{OpenAI_GPT4o}, \texttt{Kimi-K2}~\cite{team2025kimi}, \texttt{DeepSeek-V3}~\cite{liu2024deepseek}, and \texttt{Qwen3-Coder}. All models are configured with a maximum output length of 32,768 tokens. Each model is evaluated four times in each problem, and the average results are reported.

\subsection{Main Result}

\begin{table}[htbp]
\centering
\footnotesize
\caption{Performance comparison between reasoning models and non-reasoning models on AetherCode v1 (\%, 2401-2505).}
\begin{tabular}{l|cccc|cc|ccc}
\toprule
\multirow{2}{*}{\textbf{Model}} & \multicolumn{4}{c|}{\textbf{Difficulty}} & \multicolumn{2}{c|}{\textbf{Year}} & \multicolumn{3}{c}{\textbf{Pass@}}\\
\cmidrule(lr){2-5} \cmidrule(lr){6-7} \cmidrule(lr){8-10}
 & Easy & Medium & Hard & Extreme & 2024 & 2025 & 1 & 2 & 4\\
\midrule
\multicolumn{9}{l}{\textit{Reasoning Models}} \\ 
\addlinespace
\midrule
o4-mini-high & 65.3 & 32.1 & 8.0 & 3.8  & 35.8 & 32.6 & $\mathbf{35.5}$ & 43.0 & 46.6 \\
Gemini-2.5-Pro & 60.1 & 28.6 & 8.5 & 2.5& 33.7 & 25.0 & $\mathbf{32.7}$ & 39.8 & 46.0  \\
Seed-1.6-thinking-0715 & 53.9 & 20.2 & 4.7 & 0 & 28.3 & 14.7 & $\mathbf{26.6}$ & 33.0 & 38.5 \\
DeepSeek-R1-0528 & 46.2 & 16.0 & 3.8 & 0 & 23.4 & 14.3 & $\mathbf{22.3}$ & 27.4 & 32.4  \\
Gemini-2.5-Flash & 42.1 & 15.2 & 2.7 & 0 & 22.0 & 8.0 &  $\mathbf{20.3}$ & 24.5 & 28.5  \\
Qwen3-235B-A22B & 37.6 & 12.4 & 1.9 & 0 & 19.1 & 7.1 & $\mathbf{17.6}$ & 21.7 & 25.2  \\
Qwen3-32B & 34.8& 10.9 & 2.7 & 0 &  17.7 & 6.7 & $\mathbf{16.3}$ & 20.4 & 23.9  \\
Qwen3-8B & 23.7 & 4.8 & 0.8 & 0 & 11.1 & 2.7 & $\mathbf{10.0}$ & 13.0 & 15.5  \\

\midrule
\multicolumn{9}{l}{\textit{Non-Reasoning Models}} \\ 
\addlinespace
\midrule
GPT-4.1 & 23.9 & 5.7 & 1.1 & 0 & 11.3 & 4.5 & $\mathbf{10.5}$ & 13.2 & 15.3  \\
Kimi-K2 & 23.1 & 4.7 & 1.0 & 0 & 10.6 & 4.0 & $\mathbf{9.8}$ & 12.2 & 14.5  \\
DeepSeek-V3-0324 & 20.8 & 4.0 & 0 & 0  & 8.9 & 5.4 & $\mathbf{8.5}$ & 10.5 & 12.3\\
Qwen3-Coder-480B-A35B & 19.7 & 2.2 & 0.6 & 0 & 8.6 & 1.8 & $\mathbf{7.7}$ & 9.9 & 11.8  \\
GPT-4o & 11.6 & 1.0 & 0.2 & 0 &  4.9 & 1.3 & $\mathbf{4.4}$ & 5.6 & 7.0  \\

\bottomrule
\end{tabular}

\label{tab:main_result}
\end{table}

Table~\ref{tab:main_result} presents a comprehensive performance evaluation of several prominent models on AetherCode. For full results, please refer to the online leaderboard. The analysis yields the following key conclusions:
\paragraph{\textbf{Significant Performance Gap between Models}} The performance of \texttt{o4-mini-high} and \texttt{Gemini-2.5-Pro} is exceptional, establishing a significant performance gap that places them in a tier of their own above other models. Furthermore, they are the only two models capable of successfully solving problems at the "Extremely Difficult" level. Across all difficulty tiers, the performance of these two models substantially surpasses that of their competitors.

\paragraph{\textbf{Reasoning Models Comprehensively Outperform Non-Reasoning Models}} As anticipated, reasoning models demonstrate markedly superior performance compared to non-reasoning models. For instance, models from the Qwen3 series, such as \texttt{Qwen3-32B}, outperform several non-reasoning models despite having fewer parameters. More notably, even with four sampling attempts (\textit{Pass@4}), the performance of non-reasoning models still falls short of that achieved by reasoning models. This phenomenon indicates that for complex tasks like coding competitions, the solution space exploration capabilities of non-reasoning models are constrained, making it difficult to find correct solutions through limited sampling. This bottleneck is particularly pronounced in weaker models.

\paragraph{\textbf{Top-Tier Models Exhibit Great Exploration Potential}} A comparison of \textit{Pass@1} and \textit{Pass@4} scores reveals that increasing the number of samples yields a more substantial performance improvement for top-tier models. For example, \texttt{o4-mini-high}'s score improved by 11.1\% (from 35.5\% to 46.6\%), whereas the weaker \texttt{ Qwen3-32B} only saw a gain of 7.6\% (from 16.3\% to 23.9\%). Particularly noteworthy is \texttt{Gemini-2.5-Pro}, which achieved a remarkable performance increase of 13.3\% (from 32.5\% to 46.0\%). This demonstrates its vast exploration potential in solving complex programming problems, enabling it to generate more diverse and high-quality solutions through multiple attempts.

\subsection{Performance Across Algorithms}

The performance comparison in Table~\ref{tab:tag_result} reveals a significant differentiation in model capabilities across various problem categories. All models, regardless of being reasoning or non-reasoning types, uniformly excel at pattern-based tasks such as ``Basic Algorithms'' and ``Strings''. However, their limitations become equally apparent when handling highly abstract problems. Most models struggle to tackle ``Computational Geometry'' and ``Tree Structures'', with the performance of \texttt{o4-mini-high} in computational geometry being a notable exception. Furthermore, the shortcomings of non-reasoning models are particularly pronounced, as their capability bottlenecks extend into domains that also demand deep logic and abstract thinking, such as ``Dynamic Programming'' and ``Mathematics''.
\section{Conclusion}
\label{sec:conclusion}

In this paper, we introduced AetherCode, a challenging, rigorously evaluated benchmark purpose-built to assess LLMs’ coding and reasoning capabilities.  AetherCode distinguishes itself by sourcing all its problems from premier global programming competitions, including OI series and ICPC series, which ensures a high degree of challenge and relevance. Furthermore, it features a comprehensive and meticulously validated suite of test cases, created through a hybrid model of automated generation and expert curation. By validating against a dataset of over 30,000 human submissions, our test suite achieves 100\% TPR and 100\% TNR on our collected solution set, guaranteeing exceptional accuracy and reliability in evaluation.

Our comprehensive evaluation of several leading-edge models on AetherCode yielded critical insights. We observed a significant performance disparity among models, with top performers like \texttt{o4-mini-high} and \texttt{Gemini-2.5-Pro} establishing a distinct upper tier. Reasoning models demonstrated a clear and consistent advantage over their non-reasoning counterparts across all difficulty levels, highlighting the crucial role of logical deduction in solving complex algorithmic problems. Overall, even the most advanced models today can only solve a small fraction of problems in AetherCode. This indicates that current LLMs still have considerable room for improvement in reasoning and coding, and there remains a significant gap compared to top human experts.

\begin{table}[htbp]
\centering
\footnotesize
\caption{Category division and detailed tag distribution of AetherCode.}
\label{tab:algorithm_categories}
\begin{tabular}{p{4cm}|p{9cm}}
\toprule
\textbf{Category} & \textbf{Tags} \\
\midrule

\textbf{Algorithm Basics} & 
Enumeration, Simulation, Recursion, Greedy, Sorting, Divide and Conquer, Binary Search, Doubling, Recurrence \\
\midrule

\textbf{Search} & 
DFS, BFS, Bidirectional Search, Heuristic Search, A*, Iterative Deepening Search, IDA*, Dancing Links \\
\midrule

\textbf{Dynamic Programming} & 
Basic DP, Memorization Search, Knapsack DP, Range DP, DP on DAGs, Tree DP, Bitmask DP, Digit DP, Plug DP, Counting DP, Dynamic DP, Probability DP, DP Optimization \\
\midrule

\textbf{Strings} & 
String Matching, String Hashing, Trie, Palindrome Automation, Prefix Function, Z-function, Automation, AC Automation, Suffix Array, Suffix Automation, Suffix Balanced Tree, Generalized Suffix Automation, Suffix Tree, Manacher's Algorithm, KMP Algorithm, Sequence Automation, Minimal Representation, Lyndon Factorization, Main-Lorentz Algorithm \\
\midrule

\textbf{Mathematics} & 
Number Theory, Linear Algebra, Linear Programming, Abstract Algebra, Probability Theory, Game Theory, Young Matrix, Inclusion-Exclusion Principle, Combinatorics, Polynomials \\
\midrule

\textbf{Data Structures} & 
Stack, Queue, Linked List, Hash Table, Disjoint Set Union, Heap, Block Structure, Monotonic Queue, ST Table, Binary Indexed Tree, Segment Tree, Balanced Tree, Binary Tree \& Balanced Tree, Block Decomposition, Persistent Data Structures, Tree-in-Tree, K-D Tree, Cartesian Tree, Huffman Tree, STL-based Data Structure \\
\midrule

\textbf{Graph Theory} & 
Matrix-Tree Theorem, Directed Acyclic Graph, Topological Sort, Minimum Spanning Tree, Minimum Diameter Spanning Tree, Minimum Tree Spanning, Connectivity, Shortest Path, 2-SAT, Difference Constraints, Hamiltonian Graph, Modular Shortest Path, Graph Coloring, Eulerian Graph, Dominating Tree, Bipartite Graph, Prüfer Sequence, Planar Graph, Chordal Graph, Network Flow, Graph Matching, Random Walk on Graphs, LGV Lemma, Strongly Connected Components \\
\midrule

\textbf{Computational Geometry} & 
Euclidean Distance, Manhattan Distance, Chebyshev Distance, Pick's Theorem, Triangulation, Convex Hull, Sweep Line, Rotating Calipers, Half-Plane Intersection, Closest Pair of Points, Random Increment Method, Reflection Transformation, Misc. CG \\
\midrule

\textbf{Common Techniques} & 
Discretization, Two Pointer Technique, Prefix Sum \& Difference, Fractional Programming, Randomization, Hanging Line Method, Binary Thinking, Pattern Recognition, Gray Code, Expression Evaluation, Construction, Properties of Bitwise Operations, Conjecture of Conclusions, Interactive Problems, Meet in Middle, Ad-hoc, Uncertainty Algorithms, Square Root Decomposition \\
\midrule

\textbf{Problems on Trees} & 
LCA, DSU on Tree, Divide and Conquer on Points, Block Decomposition on Tree, Heavy-Light Decomposition, Chain Decomposition, Tree Diameter and Centroid, LCT \\

\bottomrule
\end{tabular}
\end{table}

\begin{table}[htbp]
\centering
\small
\caption{Performance comparison (Pass@1) between reasoning models and non-reasoning models across different categories of algorithmic problems. The abbreviations Basic, Search, DP, Str., Math, DS, Graph, Geo., Tech., Tree represent Algorithm Basics, Search, Dynamic Programming, Strings, Mathematics, Data Structures, Graph Theory, Computational Geometry, Common Techniques, and Problems on Trees, respectively.
}
\label{tab:tag_result}
\begin{tabular}{p{3.7cm}cccccccccc}
\toprule
\textbf{Model} & \textbf{Basic} & \textbf{Search} & \textbf{DP} & \textbf{Str.} & \textbf{Math} & \textbf{DS} & \textbf{Graph} & \textbf{Geo.} & \textbf{Tech.} & \textbf{Tree} \\
\midrule
\multicolumn{11}{l}{\textit{Reasoning Models}} \\ 
\addlinespace
\midrule
o4-mini-high & 38.1 & 28.5 & 27.7 & 35.6 & 31.8 & 25.8 & 28.5 & 27.1 & 26.9 & 7.3 \\
Gemini-2.5-Pro & 36.1 & 24.5 & 24.6 & 29.8 & 31.5 & 25.4 & 26.2 & 18.1 & 23.0 & 7.3 \\
Seed-1.6-thinking & 32.2 & 17.0 & 17.3 & 26.0 & 24.2 & 17.9 & 18.8 & 12.5 & 19.2 & 1.0 \\
DeepSeek-R1-0528 & 26.3 & 16.0 & 14.6 & 23.1 & 19.3 & 16.3 & 15.6 & 10.4 & 13.8 & 7.3 \\
Gemini-2.5-Flash & 24.1 & 16.5 & 11.8 & 19.2 & 16.7 & 16.3 & 17.2 & 13.2 & 11.4 & 4.2 \\
Qwen3-235B-A22B & 22.2 & 13.0 & 8.4 & 20.2 & 13.5 & 11.0 & 12.5 & 11.1 & 9.4 & 4.2 \\
Qwen3-32b & 19.7 & 11.5 & 10.9 & 18.3 & 14.1 & 11.0 & 9.4 & 6.9 & 11.2 & 0 \\

Qwen3-8B & 13.3 & 9.0 & 3.9 & 15.4 & 7.6 & 7.9 & 6.3 & 1.4 & 4.9 & 1.0 \\

\midrule
\multicolumn{9}{l}{\textit{Non-Reasoning Models}} \\ 
\addlinespace
\midrule
GPT-4.1 & 13.9 & 9.5 & 3.4 & 19.2 & 4.2 & 8.3 & 5.5 & 6.3 & 6.0 & 0 \\
Kimi-K2 & 13.7 & 7.5 & 3.6 & 15.4 & 7.0 & 8.1 & 6.6 & 0.7 & 3.6 & 0 \\
DeepSeek-V3-0324 & 12.1 & 7.0 & 1.8 & 14.4 & 3.9 & 6.3 & 4.3 & 0 & 3.6 & 0 \\
Qwen3-Coder-480B-A35B &11.1 & 5.5 & 1.8 & 14.4 & 4.2 & 5.2 & 4.3 & 1.4 & 2.9 & 1.0\\
GPT-4o & 7.2 & 4.5 & 0.7 & 11.5 & 1.6 & 2.9 & 0.4 & 0 & 1.5 & 0 \\
\bottomrule
\end{tabular}
\end{table}

\newpage
{
\scriptsize
\begin{longtable}{p{8cm} p{5cm} p{1.2cm}}
\caption{Curated Contest Source of AetherCode v1 (2401-2505).} \label{tab:datasource} \\
\toprule
\textbf{Competition Name} & \textbf{Category} & \textbf{Date} \\
\midrule
\endfirsthead

\toprule
\textbf{Competition Name} & \textbf{Category} & \textbf{Date} \\
\midrule
\endhead

\bottomrule
\endfoot

\bottomrule
\endlastfoot

Croatian Open Competition in Informatics 2023/2024 Contest \#3 & Croatian OI & 2024/1/13 \\
USACO 2024 January Contest (Platinum) & USACO Platinum & 2024/1/26 \\
The 2023-2024 ICPC Southwestern Europe Regional Contest & ICPC Regional Contests & 2024/1/28 \\
Croatian Open Competition in Informatics 2023/2024 Contest \#4 & Croatian OI & 2024/2/10 \\
USACO 2024 February Contest (Platinum) & USACO Platinum & 2024/2/16 \\
USACO 2024 US Open Contest (Platinum) & USACO Platinum & 2024/3/15 \\
Singapore National Olympiad in Informatics 2024 Final Contest & NOI (SG) & 2024/3/16 \\
Croatian Open Competition in Informatics 2023/2024 Contest \#5 & Croatian OI & 2024/3/16 \\
The 2024 ICPC Latin America Championship & ICPC Regional Championships/Finals & 2024/3/17 \\
The 2024 ICPC Europe Championship & ICPC Regional Championships/Finals & 2024/3/24 \\
The 2024 British Informatics Olympiad Final & British OI & 2024/4/6 \\
Baltic Olympiad in Informatics 2024 Day 1 & Baltic OI & 2024/5/4 \\
Baltic Olympiad in Informatics 2024 Day 2 & Baltic OI & 2024/5/5 \\
Asia-Pacific Informatics Olympiad 2024 (APIO 2024) & APIO & 2024/5/18 \\
The 2024 ICPC North America Championship & ICPC Regional Championships/Finals & 2024/5/27 \\
Central European Olympiad in Informatics 2024 Day 1 (CEOI 2024 Day 1) & Central European OI & 2024/6/25 \\
Central European Olympiad in Informatics 2024 Day 2 (CEOI 2024 Day 2) & Central European OI & 2024/6/27 \\
China National Olympiad in Informatics 2024 Day 1 & NOI & 2024/7/18 \\
China National Olympiad in Informatics 2024 Day 2 & NOI & 2024/7/20 \\
European Girls' Olympiad in Informatics 2024 Day 1 & European Girl's OI & 2024/7/23 \\
European Girls' Olympiad in Informatics 2024 Day 2 & European Girl's OI & 2024/7/25 \\
International Olympiad in Informatics 2024 Day 1 & IOI & 2024/9/3 \\
International Olympiad in Informatics 2024 Day 2 & IOI & 2024/9/5 \\
The 2024 ICPC World Finals Astana & ICPC World Finals & 2024/9/19 \\
The 2024 ICPC Kunming Invitational Contest & ICPC Regional Contests & 2024/9/28 \\
The 2024 Nordic Collegiate Programming Contest & NCPC & 2024/10/5 \\
Croatian Open Competition in Informatics 2024/2025 Contest \#1 & Croatian OI & 2024/10/5 \\
CCPC 2024 Harbin Site & CCPC & 2024/10/26 \\
The 2024 ICPC Asia Chengdu Regional Contest & ICPC Regional Contests & 2024/10/27 \\
The 2024 ICPC Asia Nanjing Regional Contest & ICPC Regional Contests & 2024/11/3 \\
Croatian Open Competition in Informatics 2024/2025 Contest \#2 & Croatian OI & 2024/11/9 \\
2024-2025 ICPC Latin American Regional Programming Contest & ICPC Regional Championships/Finals & 2024/11/9 \\
2024 Rocky Mountain Regional Contest & ICPC Regional Contests & 2024/11/9 \\
2024 North Central NA Regional Contest & ICPC Regional Contests & 2024/11/9 \\
2024 Mid-Central USA Programming Contest & ICPC Regional Contests & 2024/11/9 \\
CCPC 2024 Chongqing Site & CCPC & 2024/11/10 \\
The 2024 ICPC Greater NY Regional Contest & ICPC Regional Contests & 2024/11/10 \\
The 2024 ICPC Asia Hangzhou Regional Contest & ICPC Regional Contests & 2024/11/10 \\
CCPC 2024 Jinan Site & CCPC & 2024/11/16 \\
The 2024 ICPC Pacific Northwest Regional Contest (Div. 1) & ICPC Regional Contests & 2024/11/16 \\
The 2024 ICPC Pacific Northwest Regional Contest (Div. 2) & ICPC Regional Contests & 2024/11/16 \\
ICPC NA South Division 2024 - Division 2 & ICPC Regional Contests & 2024/11/16 \\
ICPC NA South Division 2024 - Division 1 & ICPC Regional Contests & 2024/11/16 \\
The 2024 ICPC Southern California Regional Contest & ICPC Regional Contests & 2024/11/16 \\
The 2024 ICPC Southeastern Europe Regional Contest (SEERC 2024) & ICPC Regional Contests & 2024/11/17 \\
The 2024 ICPC Asia Shanghai Regional Contest & ICPC Regional Contests & 2024/11/17 \\
The 2024 ICPC Asia Seoul Regional Contest & ICPC Regional Contests & 2024/11/23 \\
The 2024 ICPC Northwestern Europe Regional Contest (NWERC 2024) & ICPC Regional Contests & 2024/11/24 \\
The 2024 ICPC Asia Shenyang Regional Contest & ICPC Regional Contests & 2024/11/24 \\
Romanian Master of Informatics 2024 Day 1 & Romanian OI & 2024/11/28 \\
Romanian Master of Informatics 2024 Day 2 & Romanian OI & 2024/11/29 \\
The 2024 ICPC Asia Kunming Regional Contest & ICPC Regional Contests & 2024/12/1 \\
Croatian Open Competition in Informatics 2024/2025 Contest \#3 & Croatian OI & 2024/12/12 \\
USACO 2024 December Contest (Platinum) & USACO Platinum & 2024/12/13 \\
The 2024 ICPC Northern Eurasia Finals & ICPC Regional Championships/Finals & 2024/12/15 \\
The 2024 ICPC Central Europe Regional Contest & ICPC Regional Contests & 2024/12/15 \\
CCPC 2024 Zhengzhou Site & CCPC & 2024/12/21 \\
The 2024 ICPC Asia Yokohama Regional Contest & ICPC Regional Contests & 2024/12/22 \\
The 2024 ICPC Asia Hong Kong Regional Contest & ICPC Regional Contests & 2024/12/22 \\
The 2024 ICPC Asia East Continent Final Contest & ICPC Regional Championships/Finals & 2024/12/28 \\
USACO 2025 January Contest (Platinum) & USACO Platinum & 2025/1/24 \\
Croatian Open Competition in Informatics 2024/2025 Contest \#4 & Croatian OI & 2025/1/25 \\
The 24th Japanese Olympiad in Informatics Final Round (JOI 2024/2025) & Japanese OI & 2025/2/2 \\
Croatian Open Competition in Informatics 2024/2025 Contest \#5 & Croatian OI & 2025/2/15 \\
USACO 2025 February Contest (Platinum) & USACO Platinum & 2025/2/21 \\
The 2025 ICPC Europe Championship & ICPC Regional Championships/Finals & 2025/3/2 \\
2025 ICPC Asia West Finals & ICPC Regional Championships/Finals & 2025/3/7 \\
The 2025 ICPC Latin America Championship & ICPC Regional Championships/Finals & 2025/3/16 \\
USACO 2025 US Open Contest (Platinum) & USACO Platinum & 2025/3/21 \\
Singapore National Olympiad in Informatics 2025 Final Contest & NOI (SG) & 2025/3/22 \\
The 2025 British Informatics Olympiad Final & British OI & 2025/4/12 \\
Baltic Olympiad in Informatics 2025 Day 1 & Baltic OI & 2025/4/26 \\
Baltic Olympiad in Informatics 2025 Day 2 & Baltic OI & 2025/4/27 \\
The 2025 ICPC China Zhejiang Province Programming Contest (22nd) & ICPC Regional Contests & 2025/5/10 \\
CCPC Final 2024 & CCPC Final & 2025/5/11 \\
Asia-Pacific Informatics Olympiad 2025 (APIO 2025) & APIO & 2025/5/17 \\
The 2025 ICPC Asia Wuhan Invitational Contest & ICPC Regional Contests & 2025/5/17 \\
The 2025 ICPC North America Championship & ICPC Regional Championships/Finals & 2025/5/26 \\
\end{longtable}
}

\newpage
\section*{Contributions}

\subsection*{Research \& Development}

Zihan Wang$^{1,2}$, Jiaze Chen$^1$, Zhicheng Liu$^1$

\subsection*{Management}

Markus Mak$^1$, Yidi Du$^1$

\subsection*{Operations}

Geonsik Moon$^1$, Luoqi Xu$^1$, Aaron Tua$^1$

\subsection*{Expert Partner}

Kunshuo Peng$^1$, Jiayi Lu$^1$, Mingfei Xia$^1$

\subsection*{Data Operations}

Boqian Zou$^1$, Chenyang Ran$^1$, Guang Tian$^1$, Shoutai Zhu$^1$, Yeheng Duan$^1$, Zhenghui Kang$^1$

\subsection*{Data Platform}

\textbf{Front-End Developer}: Zhenxing Lin$^1$, Shangshu Li$^1$

\textbf{Back-End Developer}: Qiang Luo$^1$, Qingshen Long$^1$

\textbf{Product Manager}: Zhiyong Chen$^1$, Yihan Xiao$^1$

\subsection*{Writing}

Yurong Wu$^1$, Daoguang Zan$^1$

\subsection*{Supervision}

Yuyi Fu$^1$, Mingxuan Wang$^1$, Ming Ding$^1$

\subsection*{Affiliations}

$^1$ByteDance

$^2$M-A-P

\section*{Acknowledgments}

We thank Siyao Liu, Jinxin Chi, Haojie Pan, Jingjing Xu, Ge Zhang, Wenhao Huang, Yonghui Wu, as well as other colleagues at ByteDance, and more importantly, the anonymized competitive programming expert team, for their support for the AetherCode project.

\newpage

\section*{Disclaimer}

Your access to and use of this dataset are at your own risk. We do not guarantee the accuracy of this dataset. The dataset is provided “as is” and we make no warranty or representation to you with respect to it and we expressly disclaim, and hereby expressly waive, all warranties, express, implied, statutory or otherwise. This includes, without limitation, warranties of quality, performance, merchantability or fitness for a particular purpose, non-infringement, absence of latent or other defects, accuracy, or the presence or absence of errors, whether or not known or discoverable. 

In no event will we be liable to you on any legal theory (including, without limitation, negligence) or otherwise for any direct, special, indirect, incidental, consequential, punitive, exemplary, or other losses, costs, expenses, or damages arising out of this public license or use of the licensed material.

The disclaimer of warranties and limitation of liability provided above shall be interpreted in a manner that, to the extent possible, most closely approximates an absolute disclaimer and waiver of all liability.

\clearpage

\bibliographystyle{plainnat}
\bibliography{main}

\clearpage



\end{document}